\documentclass[11pt]{article}
 \usepackage{jheppub}

\usepackage{color}
\usepackage{amsmath}
\usepackage{verbatim}
\usepackage{subfigure}
\usepackage{acronym}

\usepackage{amsfonts}
\usepackage{amssymb}
\usepackage{mathrsfs}
\usepackage{graphicx}
\usepackage{multirow}
 \usepackage{slashed}
 \usepackage{epsfig,multicol,bbm}
 \usepackage{url}

 \usepackage[compat=1.1.0]{tikz-feynman}

\definecolor{commam}{rgb}{0.2,0.5,1.0}
\definecolor{myred}{rgb}{0.9, 0.5, 0.1}
\definecolor{myblue}{rgb}{0, 0, 0.7}
\definecolor{mygreen}{rgb}{0.04, 0.7, 0.5}
\definecolor{orange}{rgb}{1,0.5,0}

\hypersetup{colorlinks,citecolor=myred,linkcolor=myblue,urlcolor=myblue,linktocpage=true}

\def\beq{\begin{equation}\begin{aligned}}
\def\eeq{\end{aligned}\end{equation}}
 \def\be   {\begin{equation}}   \def\ee   {\end{equation}}
 \def\ba   {\begin{array}}      \def\ea   {\end{array}}
 \def\bea  {\begin{eqnarray}}   \def\eea  {\end{eqnarray}}
 \def\bean {\begin{eqnarray*}}  \def\eean {\end{eqnarray*}}
 
 \def\bry{\begin{array}}
 \def\ery{\end{array}}

\begin{document}

\title{Electroweak Symmetry Non-Restoration\\ from Dark Matter} 

\author[a]{Oleksii Matsedonskyi,}

\author[b]{James Unwin,}

\author[b]{and Qingyun Wang}

\affiliation[a]{Weizmann Institute of Science, 234 Herzl Street, Rehovot 7610001, Israel}

\affiliation[b]{Department of Physics, University of Illinois at Chicago, Chicago, IL 60607, USA}

\date{\today}

\abstract{
Restoration of the electroweak symmetry at temperatures around the Higgs mass is linked to tight phenomenological constraints on many baryogenesis scenarios. A potential remedy can be found in mechanisms of electroweak symmetry non-restoration (SNR), in which symmetry breaking is extended to higher temperatures due to new states with couplings to the Standard Model. Here we show that, in the presence of a second Higgs doublet, SNR can be realized with only a handful of new fermions which can be identified as viable dark matter candidates consistent with all current observational constraints. The competing requirements on this class of models allow for SNR at temperatures up to $\sim$TeV, and imply the presence of sub-TeV new physics with sizable interactions with the Standard Model.  As a result this scenario is highly testable with signals in reach of next-generation collider and dark matter direct detection experiments.}

\maketitle


\section{Introduction}

Given the form of the Higgs potential and the particle contents of the Standard Model (SM), it is a common expectation that at temperatures around $160$~GeV the electroweak symmetry should be restored to an unbroken gauge symmetry. However, if the Higgs potential is appropriately modified (or some other source of electroweak symmetry breaking is added) then the electroweak symmetry can remain unbroken to higher temperatures. 
The general concept of symmetry non-restoration (SNR) was first explored by Weinberg \cite{Weinberg:1974hy} and then expanded upon in \cite{Mohapatra:1979qt,Fujimoto:1984hr,Dvali:1995cj,Salomonson:1984rh,Bimonte:1995sc,Dvali:1996zr,Kilic:2015joa,Orloff:1996yn,Gavela:1998ux,Ahriche:2010kh,Espinosa:2004pn,Bajc:1998jr,Agrawal:2021alq}. Recently, scenarios of SNR for the electroweak sector, and the associated phenomenological observables, have become a topic of investigation \cite{Meade:2018saz,Baldes:2018nel,Glioti:2018roy,Matsedonskyi:2020mlz,Matsedonskyi:2020kuy,Carena:2021onl,Biekotter:2021ysx,Bai:2021hfb}. Electroweak SNR is of special interest in the context of certain baryogenesis scenarios \cite{Cline:2006ts}. 
In particular, electroweak SNR can change the minimal temperature at which sphalerons are active if the scale of electroweak restoration is altered \cite{Baldes:2018nel,Glioti:2018roy}. 
For instance, this allows the realization of electroweak baryogenesis at higher temperatures, with a corresponding increase of the new physics scales involved, and a suppression of certain unwanted new physics effects such as contributions to electron EDM.
Moreover, if the temperature of the universe following inflationary reheating is lower than the scale of electroweak symmetry restoration, then sphaleron effects can be entirely eliminated thus prohibiting sphalerons from transferring or washing-out of asymmetries generated in the early universe.

\newpage

One of the defining features of electroweak SNR is the presence of new states interacting with the SM, which are abundant in the early universe and can {\it a priori} be stable on cosmological time scales. However, within the previously proposed scenarios, these new states cannot serve as viable dark matter (DM) candidates. 
We show that this issue can be resolved by extending the scenario with a second Higgs doublet\footnote{More generally, one could use any $SU(2)$ multiplet for this purpose.}. Besides linking SNR with DM, such an extension allows one to significantly decrease the number of new degrees of freedom required for SNR~{\cite{Glioti:2018roy,Carena:2021onl}}; notably, the original one-Higgs SNR mechanisms required at least ${\cal O}(10)$ new Dirac fermions, or at least ${\cal O}(100)$ new singlet scalars.

In this work we will concentrate on SNR induced by new singlet fermions  \cite{Matsedonskyi:2020mlz}.  Compared to the case of singlet scalars  \cite{Meade:2018saz,Baldes:2018nel,Glioti:2018roy}, this class of SNR allows for a significant reduction in the overall number of new degrees of freedom, at the price of lower maximal temperatures where it is effective. The number of required new fermions grows with the square of the maximal temperature and starts being a disadvantage compared to the scalar case for temperatures in excess of a few TeV. This limitation however stops being relevant for instance when electroweak baryogenesis is considered in the context of such gauge hierarchy problem motivated scenarios as composite Higgs models~\cite{Espinosa:2011eu,Chala:2016ykx,Bruggisser:2018mus,Bruggisser:2018mrt} or the relaxion mechanism~\cite{Graham:2015cka}. In the former, the Higgs field ``dissolves'' at temperatures above $\sim$TeV, while in the latter the reheat temperatures after inflation are often constrained to be similarly low. Therefore electroweak baryogenesis in both cases has to operate at $T\lesssim 1$~TeV even in the presence of SNR.
Moreover, it was shown in~\cite{Matsedonskyi:2020kuy} that in the specific case of composite Twin Higgs models the new states triggering SNR can be naturally present in the spectrum of the model. It might also be highlighted that, generally speaking, embedding of scalar-driven SNR in a naturalness-motivated scenario would at least require a reconsideration due to the presence of new physics needed to explain the lightness of the Higgs and the multitude of SNR scalars.

Notably, here we show that with the inclusion of the second Higgs doublet the number of new states required for SNR can be as little as a single pair of new fermions. We also find that the competing requirements of SNR models with DM tend to force the viable parameter space to be relatively constrained, making these scenarios eminently testable. Moreover, the model constraints imply that a relatively low mass for the second Higgs is a general feature, with a sizable coupling between the two Higgs doublets, leading to an abundance of complementary phenomenological signals at next-generation collider and direct detection experiments. 

This paper is structured as follows. We start by exposing the difficulties in connecting the SNR states with DM for the case of models with a single (SM) Higgs doublet in Section~\ref{sec:1hdm}. Subsequent, in Section~\ref{sec:2hdm} we introduce SNR in the context of two Higgs doublet models. Identifying the new fermion states which induce SNR with DM, in Section~\ref{sec:expbounds} we examine the compatibility of this class of models with the current experimental bounds. Section~\ref{sec:remarks} contains an overview of our results and some concluding remarks.

\section{SNR and DM}\label{sec:1hdm}

We shall start by making some comments regarding the need for a second Higgs doublet in scenarios in which SNR is linked to the DM relic density. Specifically, we shall highlight the issues that arise in models in which the SM Higgs alone is the only significant source of electroweak symmetry breaking.

We will consider scenarios where the electroweak symmetry is broken starting from temperatures of the order 1~TeV and remains broken all the way down to $T=0$, while maintaining $h/T>1$, which is a critical condition to ensure that the baryon asymmetry is not washed out by electroweak sphalerons  \cite{Quiros:1999jp}. For $T\lesssim 130$~GeV a sufficient amount of electroweak symmetry breaking is produced by the SM zero-temperature Higgs potential. However,  to ensure that $h/T$ stays greater than unity at temperatures above $130$ GeV one needs to compensate the positive Higgs thermal mass induced by the plasma of SM particles which acts to drive the Higgs vacuum expectation value (VEV) to zero
(we collect the relevant expressions for the thermal corrections in Appendix~\ref{app:tcorr})
\beq\label{eq:smTmass}
\delta V_h(T) \simeq \frac 1 2 \left[
\frac{\lambda_t^2}{4} +\frac{\lambda}{2} + \frac{3g^2}{16} + \frac{g^{\prime 2}}{16} 
\right] h^2  \simeq \frac{0.4}{2}\, T^2 h^2~,  
\eeq
where $\lambda_t$ is the top Yukawa coupling, $\lambda$ is the Higgs quartic, $g$ and $g'$ are the electroweak couplings, and $h$ is the physical Higgs boson.  We will focus on the case in which high-temperature SNR is achieved by adding $n_\chi$ new fermions\footnote{The essential feature is the multiplicity factor and thus this could equally be realised as a single fermion state transforming under some global symmetry with $n_\chi$ degrees of freedom.} with the Higgs-dependent mass term
\beq
{\cal L} = -m_{\chi0} \overline \chi \chi + \frac{1}{\Lambda} h^2 \overline \chi \chi\,,
\eeq
which induces the following thermal correction to the Higgs potential 
\beq\label{eq:VofTh1}
\delta V_h(T)_{\text{SNR}}\simeq n_\chi \frac{T^2 m_\chi^2[h]}{12} \supset -  \frac {n_\chi} {6}\frac{m_{\chi0}}{\Lambda} T^2 h^2 + \frac {n_\chi} {12} \frac{T^2}{\Lambda^2} h^4.  
\eeq
The negative mass correction contained in Eq.~(\ref{eq:VofTh1}) dominates over the symmetry-restoring effect of the SM states in Eq.~(\ref{eq:smTmass}) provided
\beq\label{eq:snr1}
n_\chi \frac {m_{\chi 0}}{\Lambda} \gtrsim 1.
\eeq
Furthermore, the perturbative expansion in the considered thermal field theory is justified up to temperatures of the order~\cite{Matsedonskyi:2020mlz} 
\beq\label{eq:pert1}
T \simeq \Lambda/\sqrt{n_\chi}.
\eeq
Combining Eqs.~(\ref{eq:snr1}) \& (\ref{eq:pert1}) one obtains the following bound on the number of new states required to maintain controllable SNR up to some temperature $T$
\beq\label{eq:nchi1hdm}
n_\chi \gtrsim T^2/m_{\chi 0}^2.
\eeq 
The estimate of Eq.~(\ref{eq:snr1}) only holds for the fermions which are sufficiently light, $m_\chi \lesssim T$, at all the relevant temperatures. The lowest such a temperature is $T\simeq 130$~GeV, below which the SM zero-temperature potential is sufficient to ensure $h/T>1$.  For the estimates in this section we will therefore assume $m_\chi \sim m_{\chi0} \sim 150$~GeV, although, as we will show later on, the optimal $\chi$ mass for SNR is in fact somewhat higher which however would not affect qualitatively the conclusions derived below. It follows from Eq.~(\ref{eq:nchi1hdm}) that at least $10 - 40$ new fermions $\chi$ are required to ensure SNR up to temperatures of $T=0.5-1$~TeV (which is the typical scale of maximal SNR temperatures that we will be interested in).
 
Let us now estimate the density of $\chi$ states today, where we will assume that the $\chi$ are stable  (this can be achieved via a $Z_2$ symmetry
$\chi\rightarrow-\chi$) and their relic abundance is set by the standard freeze-out mechanism controlled by the $\overline \chi \chi |H|^2$ interactions. The $\chi$ states dominant annihilation channels are $\chi \chi \to hh, W^+W^-, ZZ$, the cross-section for which is parametrically
\beq
\sigma v&\simeq \frac{v^2}{8\pi\Lambda^{2}},
\eeq
and it follows that the $\chi$ relic density is given by
\beq\label{eq:relic1hs}
\Omega_\chi h^{2}
\simeq
0.12 \left(\frac{\sqrt{n_\chi} \Lambda}{\,1.1 \,\text{TeV}} \right)^2.
\eeq
More precise analytic forms for the cross section and relic abundance are given in Appendix~\ref{app:relic}.
Observe from Eq.~(\ref{eq:relic1hs}) that even with just $n_\chi=10$ new fermions, the correct DM relic density can only be reproduced at the price of an unacceptably low cutoff: $\Lambda \simeq 350$~GeV. Moreover, perturbativity is limited to the temperatures which are even lower, $T\lesssim \Lambda/\sqrt{n_\chi}$, see Eq.~(\ref{eq:nchi1hdm}). One of the reasons for this outcome is the $p$-wave suppression of the annihilation cross-section. This can be remedied by adding a pseudo-scalar interaction  $\overline \chi i\gamma^{5}\chi h^{2}/\tilde\Lambda$ which provides an s-wave annihilation channel, permitting the correct $\chi$ relic density to be reproduced with higher energy cutoff scales:
\beq\label{eq:relic1hps}
\Omega_\chi h^{2}
\simeq
0.12 \left(\frac{\sqrt{n_\chi} \tilde\Lambda}{\,6.3 \,\text{TeV}} \right)^2.
\eeq
For $n_\chi=10$ the scale of pseudoscalar interactions is thus fixed to a more moderate value of $\tilde \Lambda\simeq 2$~TeV. Therefore, henceforth we will consider models in which $\chi$ has both a scalar and psuedoscalar coupling to the Higgs and while $\tilde\Lambda$ is fixed by the DM relic abundance condition, the scalar interaction scale $\Lambda$ is unconstrained at this stage.

The second major concern with linking $\chi$ to the DM are the experimental constraints from DM direct detection. The spin-independent DM-nucleon cross-section induced by the scalar interaction (see Appendix~\ref{app:dd}) is currently bounded as follows~\cite{Aprile:2018dbl}
\beq
\sigma_{\text{SI}} \simeq  2.8 \times 10^{-43}\,\text{cm}^2 \left( 1\,\text{TeV}/\Lambda \right)^2 \lesssim 1.1 \times 10^{-46}\,\text{cm}^2
\eeq
thereby imposing the constraint
\beq
\Lambda \gtrsim 50\,\text{TeV}.
\eeq
Such a large suppression of Higgs-$\chi$ interactions then has to be compensated by a large number of new fermions in order to have SNR, see Eq.~(\ref{eq:snr1}), $n_\chi > \Lambda/m_{\chi 0} \gtrsim 300$. This number is in turn incompatible with the relic density requirement of Eq.~(\ref{eq:relic1hps}) unless $\tilde \Lambda\lesssim 350$~GeV. 
At the same time the maximal temperature at which the effective field theory (EFT) is valid in the presence of the non-renormalizable pseudo-scalar operator is constrained to be $T<\tilde \Lambda/\sqrt{n_\chi}$, analogously to Eq.~(\ref{eq:pert1}), which again confines the validity of the theory to unacceptably small temperatures. Moreover, we find that even the special case of resonant DM annihilation (with $m_h\approx2m_\chi$) is excluded due to the limits on Higgs physics. 

We conclude that in the minimal models with a single SM Higgs boson presented here, it is  not possible to identify the SNR fermions as viable DM candidates. The tension between direct detection and the relic abundance was also found in the models with SNR induced by singlet scalars~\cite{Baldes:2018nel,Glioti:2018roy}, which do not rely on non-renormalizable interactions. The authors of~\cite{Glioti:2018roy} identified some variants in which DM could be included in the model, however the DM candidate was largely decoupled from the SNR mechanism. In this work we specifically restrict our attention to the case that the fields which induce electroweak SNR are at the same time viable DM candidates. In order to allow for that we will separate the SNR sector from the SM quark sector by introducing the second Higgs doublet, as we will discuss now.

\section{SNR with a Second Higgs Doublet}\label{sec:2hdm}

\begin{figure}[t]
\begin{center}
\hspace{-1.0cm}
\includegraphics[width=185pt]{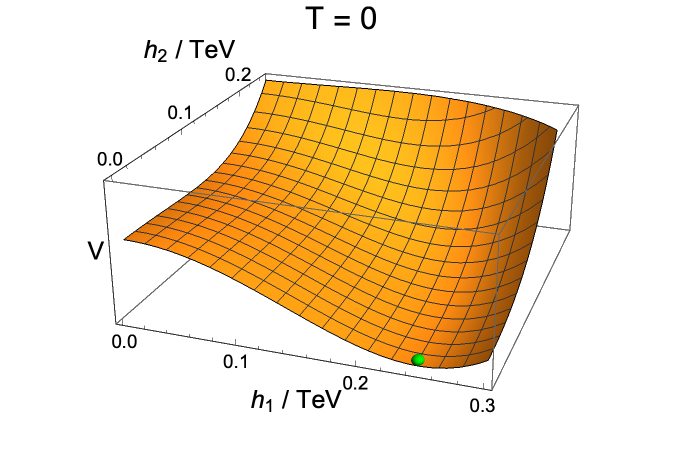}
\hspace{-1.2cm}
\includegraphics[width=185pt]{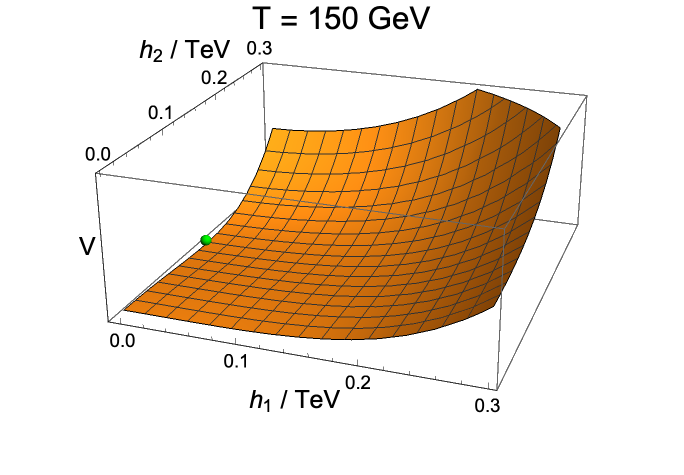} 
\hspace{-0.5cm}
\includegraphics[width=120pt]{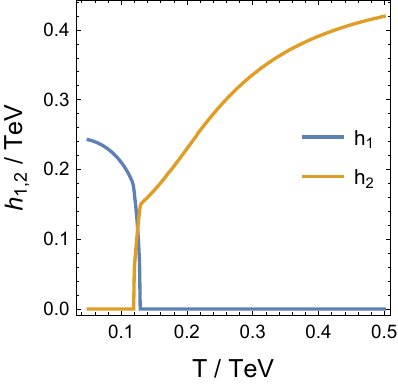}
\end{center}
\vspace{-8mm}
\caption{\small \it{Example of scalar potential at $T=0$ (left) and $T=150$~GeV (center), green points show the minima. The right panel shows the position of the global minimum in terms of $h_1$ and $h_2$ as a function of temperature. The parameters are set as follows: $n_\chi=6$, $m_{\chi 0}=0.3$~TeV, $m_{h2}=m_{h1}/2$, $\lambda_{2}=0.1$, $\lambda_{12}=0.25$, $\Lambda=1$~TeV, $\tilde \Lambda=6$~TeV, and $\mu=1$~TeV.}\label{fig:2hdm_example}}
\end{figure}

We now consider a modified scenario in which electroweak symmetry breaking at high temperature is due to a second Higgs doublet.
In Figure~\ref{fig:2hdm_example} we show an example of the thermal evolution of the Higgs VEVs, which we will analyze in detail in the remainder  of this section.
The parts of the Lagrangian relevant for electroweak symmetry breaking are
\beq
\label{eq:v2hdmtree}
	-\mathcal{L}\supset& - m_{1}^2|H_1|^2+\lambda_{1}|H_1|^4+ m_{2}^{2}|H_2|^2+\lambda_{2} |H_2|^4+\lambda_{12} |H_1|^2 |H_2|^2 \\
	&+m_{\chi{0}}\overline\chi\chi- 2\,\overline\chi\left[\frac{1}{\Lambda}+\frac{i\gamma^{5}}{\tilde\Lambda}\right]\chi \, |H_2|^2.
\eeq
In the above Lagrangian we assume no interactions of the following types (we omit h.c.~counterparts) $H_1^\dagger H_2$, $H_1^\dagger H_2 |H_1|^2$,  $H_1^\dagger H_2 |H_2|^2$  $(H_1^\dagger H_2)^2$, $H_1^\dagger H_2 H_2^\dagger H_1$, these can be forbidden by $H_2$ transforming under a combination of $Z_2$ and global $U(1)$ symmetries.\footnote{Specifically, $H_1^\dagger H_2$, $H_1^\dagger H_2 |H_1|^2$,  $H_1^\dagger H_2 |H_2|^2$ is forbidden by imposing a $Z_2$ symmetry acting on $H_2$. $(H_1^\dagger H_2)^2$ is forbidden by $U(1)_{H2}$ acting on $H_2$~\cite{Carena:2021onl}. $H_1^\dagger H_2 H_2^\dagger H_1$ is related to the previous one by a custodial symmetry (weakly broken at one-loop level), the term hence vanishes if the custodial symmetry is imposed together with $U(1)_{H2}$~\cite{Pomarol:1993mu,Carena:2021onl}.}
A small coupling between $H_1$ and $\chi$ is induced at one-loop level, which we will account for when analyzing the DM direct detection constraints. 
All the parameters of Eq.~(\ref{eq:v2hdmtree}) are assumed to be real.

 The crucial assumptions which allow this scenario to circumvent the issues encountered in Section \ref{sec:1hdm} are that we insist that second Higgs doublet has negligible couplings to SM quarks\footnote{This coupling structure also avoids dangerous tree-level flavor changing neutral currents, similar to the classic Type I two Higgs doublet model \cite{Glashow:1976nt}.} and modest couplings to the new SNR inducing fermions $\chi$. The first assumption decreases the size of the positive thermal mass, facilitating SNR, and at the same time suppresses the direct detection cross-section once the $\chi$ states are identified as DM. 
Furthermore, in order to comply with the current Higgs couplings measurements and eliminate tree-level direct detection processes we require that at zero temperature the second Higgs doublet has a vanishing VEV, such that all the electroweak symmetry breaking is induced by the first, SM-like Higgs doublet.  
While postulating suppressed couplings to quarks, we will still assume the presence of small Yukawa couplings between the second Higgs and the SM fermions, allowing the former to decay, and hence not to contribute to the DM density.

Having in mind models of electroweak baryogenesis as the main motivation, we will now analyze how to ensure the condition
\beq\label{eqsig}
h_\Sigma \equiv (h_1^2+h_2^2)^{1/2} > T
\eeq
for all temperatures below some $T\sim 0.5...1$~TeV. Provided that the condition Eq.~(\ref{eqsig}) is satisfied this will preserve any baryon asymmetry from sphaleron washout.

\subsection{Vacuum Structure}

We will first analyze the global structure of the zero-temperature potential. The physical masses of the Higgs boson and four massive components of $H_2$ in the desired global minimum at $\{h_1,h_2\}=\{v_{\text{SM}},0\}$ are given, at tree level, respectively by
\begin{eqnarray}
m_{h1}^2 &=& 2\, m_1^2\,,\\
m_{h2}^2 &=& m_2^2 + \lambda_{12}\, v_{\text{SM}}^2/2\,. \label{eq:mh2} 
\end{eqnarray}
Besides taking $\lambda_{1}, \lambda_{2}>0$, we will also assume $\lambda_{12}>0$, which will be required for SNR as we will see shortly. These assumptions ensure the absence of run-away directions at tree level.
For  $m_2^2>0$ the potential has only one minimum, which is the correct Standard Model electroweak symmetry breaking vacuum: $\{h_1,h_2\}=\{v_{\text{SM}},0\}$. 
However for $m_2^2<0$ the scalar potential can feature a second minimum, which can be deeper or shallower.

The case of a deeper minimum would make the Standard Model a valse vacuum, which can be phenomenologically problematic \cite{Barroso:2013awa}.
At tree level, a second deeper minimum develops if $m_2^2<0$, and $m_2^4/\lambda_2>m_1^4/\lambda_1$, thus to avoid such situations we require that
\beq\label{eq:correctglobalmin}
\text{\it correct global minimum:}\quad
\frac {m_1^4}{\lambda_1} > \frac {m_2^4}{\lambda_2}\,.
\eeq
On the other hand, there is a phenomenologically viable possibility of having a second minimum with $h_1=0, h_2\ne0$ which is shallower than the global minimum at $\{h_1,h_2\}=\{v_{\text{SM}},0\}$. Existence of the shallower second minimum requires, besides satisfying $m_2^2<0$ and Eq.~(\ref{eq:correctglobalmin}), also the condition $V''_{h_1}>0$ at $h_2^2 = {|m_2^2|}/{\lambda_2}$, which leads to the constraint
\beq\label{eq:secondlocmin}
\text{\it second local minimum:}\quad
\lambda_{12} |m_{2}|^2 > 2\lambda_2 m_1^2.
\eeq 

In summary, condition (\ref{eq:correctglobalmin}) ensures that the global minimum reproduces the Standard Model, but contains no restriction on the presence of possible additional metastable minima. While condition (\ref{eq:secondlocmin}) corresponds to the existence of a second (global or false) minimum at $h_1=0$ and $h_2\neq0$. Therefore, these two conditions combined ensure that the second minimum at $h_1=0$ and $h_2\neq0$ is only a local minimum with the global minimum at $\{h_1,h_2\}=\{v_{\text{SM}},0\}$.
Moreover, taking the correct global minimum condition (\ref{eq:correctglobalmin}) along with the converse of condition (\ref{eq:secondlocmin}) --by flipping the inequality-- identifies parameter points with only a single minimum at $\{h_1,h_2\}=\{v_{\text{SM}},0\}$ for $m_2^2<0$.
These tree-level conditions will be used below to derive analytic estimates for SNR, while in our parameter space scans we will use one-loop zero-temperature potential (see Appendix~\ref{app:veff}) and check the presence of the extra minima numerically.

\subsection{$H_{1,2}$ Thermal Potential}

In the early universe the evolution of the Higgs fields is governed by the thermal potential whose approximate form in the high-temperature expansion is
\begin{align}\label{Vfull}
	V(T)
	\;\supset& 
	\frac{1}{2} \left[-m_1^2+T^2 \left(c_{T1}+\frac{\lambda_{12}}{6}\right)\right]h_1^2+\frac14\lambda_1 h_1^4 \nonumber\\
	&+\;
	\frac12 \left[m_2^2+T^2 \left(c_{T2}+\frac{\lambda_{12}}{6}+\frac{\lambda_{2}}{2}-\frac{n_{\chi}m_{\chi0}}{3\Lambda}\right)\right]h_{2}^{2}+\frac14 \left[\lambda_2+\frac{n_\chi T^2}{3\Lambda^2} \left(1+\frac{\Lambda^{2}}{\tilde\Lambda^2}\right)\right]h_{2}^{4} \nonumber\\
	&+\; \frac14 \lambda_{12} h_1^2 h_2^2\,,
\end{align}
where
\begin{eqnarray}
c_{T1} &=& \frac{\lambda_t^2}{4} +\frac{\lambda_{1}}{2} + \frac{3g^2}{16} + \frac{g^{\prime 2}}{16} \simeq 0.4 \,,\\
c_{T2} &=& \frac{3g^2}{16} + \frac{g^{\prime 2}}{16} \simeq 0.1\,, \label{eq:ct2}
\end{eqnarray}
incorporate the fixed contributions proportional to electroweak gauge couplings, top Yukawa coupling and $H_1$ self-quartic.
Here $h_{1,2}$ denote the $H_{1,2}$ components taking a VEV, whose directions in $SU(2)_L$ we assume to be aligned (though in most cases this is unimportant). 

Notice the following feature of the contribution of new fermions to Eq.~(\ref{Vfull}): while the overall correction to the Higgs mass in the high-$T$ expansion is $\delta m_{h2}^2 \simeq n_\chi m_{\chi{0}} T^2 /3 \Lambda$ and formally grows with $m_{\chi{0}}$, this only holds as long as $m_{\chi {0}} \lesssim T$, where the high-$T$ expansion is valid. Otherwise the correction becomes exponentially suppressed, which reflects the depletion of the $\chi$ density in plasma. 
The optimal balance between the linear growth and the suppression is achieved for $(m_{\chi0}/T)^2\sim 2$, where the thermal correction to the mass reaches its maximal absolute value for a given $T$. The size of the correction at this point 
is about a half of what would be predicted from the high-$T$ expansion for given values of $m_{\chi0}$ and $T$ (see Appendix~\ref{app:tcorr}). Note that we are only using the high-$T$ expansion for the sake of gaining analytic understanding, and a full one-loop thermal potential (see Appendix~\ref{app:tcorr}) is used in the scans presented in Section~\ref{sec:numscan}.

\subsection{Conditions for SNR} \label{s3.3}

Using the analytic approximation Eq.~(\ref{Vfull}) we will now discuss the conditions necessary to maintain $h_\Sigma/T>1$
from temperatures of at least $T\sim 0.5$~TeV down to $T=0$, and then move to the numerical results in Section~\ref{sec:numscan}.
At very high temperatures the evolution of the first Higgs doublet is dominated by the SM thermal corrections $\propto c_{T1}$, which set $h_1=0$ analogously to what happens in the SM. The VEV of the second Higgs at high temperature has the following form
\beq\label{eq:h2vev}
h_2^2 = - m_{h2}^2(T)/\lambda_{2}(T)\,,
\eeq
where $m_{h2}^2(T)$ and $\lambda_{2}(T)$ are effective mass and quartic coupling of $h_2$ which can be read from the second line of Eq.~(\ref{Vfull}) and are given by
\begin{eqnarray}
 m_{h2}^2(T) &=& m_2^2+T^2 \left(c_{T2}+\frac{\lambda_{12}}{6}+\frac{\lambda_{2}}{2}-\frac{n_{\chi}m_{\chi0}}{3\Lambda}\right), \\
 \lambda_{2}(T) &=& \lambda_2+\frac{n_\chi T^2}{3\Lambda^2} \left(1+\frac{\Lambda^{2}}{\tilde\Lambda^2}\right).
\end{eqnarray}

It is useful to write the requirement $h_\Sigma=h_2>T$, using the expression for the $h_2$ VEV in Eq.~(\ref{eq:h2vev}), as a bound on the combination of the number of new fermions $n_\chi $, their mass $m_{\chi 0}$ and coupling strength $1/\Lambda$. Thus we rewrite $h_2>T$ as follows
\beq\label{eq:alphasnr}
\alpha_{\text{SNR}} \equiv n_\chi \frac{m_{\chi 0}}{\Lambda} \, > \,
3 \frac{m_{h2}^2-\lambda_{12} v_{\text{SM}}^2/2}{T^2} +3\, c_{T2}+ \frac 1 2 (\lambda_{12}+9 \lambda_2)  + {n_\chi} \left(\frac{T^2}{\Lambda^2}+\frac{T^2}{\tilde\Lambda^2}\right).
\eeq

At very high temperatures, the bound on $\alpha_{\text{SNR}}$ is mostly determined by the last terms of Eq.~(\ref{eq:alphasnr}) which are proportional to $T^2$ and originate from the thermal correction to the quartic term of the $h_2$ potential. From inspection of Eq.~(\ref{eq:alphasnr}) we can derive the maximal $T$ satisfying Eq.~(\ref{eq:alphasnr}), to obtain
\beq\label{eq:tsnrmax1}
T < \sqrt{m_{\chi 0} \Lambda}\, \left(1+\frac{\Lambda^2}{\tilde \Lambda^2}\right)^{-1/2}\,.
\eeq
Another upper bound on the SNR temperature comes from the validity of perturbative expansion in our finite-temperature EFT~\cite{Matsedonskyi:2020mlz} 
\beq \label{eq:pert}
T \lesssim \frac{\text{min}[\Lambda,\tilde\Lambda]}{\sqrt{n_\chi}}\,.
\eeq
In our numerical study we find that both upper bounds on the temperature are important, depending on $n_\chi$.

At lower temperatures, the right-hand side of Eq.~(\ref{eq:alphasnr}) is dominated first by the constant terms and then by the terms $\propto 1/T^2$. 
Notice the presence of the negative contribution $\propto - \lambda_{12}/T^2$, this can be understood from the fact that $\lambda_{12}$ controls the change of the second Higgs mass between $h_1=v_{\text{SM}}$ minimum and $h_1=0$, see Eq.~(\ref{eq:mh2}). For a fixed physical mass of the second Higgs at $h_1=v_{\text{SM}}$, its mass at $h_1=0$ decreases if  $\lambda_{12}$ is positive, which facilitates SNR at low temperatures.

\begin{figure}[t]
\begin{center}
\includegraphics[width=240pt]{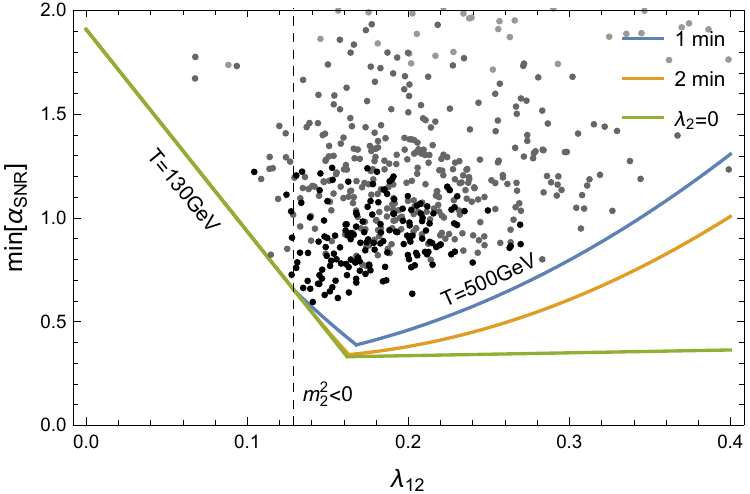}
\end{center}
\caption{\small \it{Estimates of the minimal $\alpha_{\rm{SNR}}$ as a function of $\lambda_{12}$. We set $m_{h2}=63$~GeV and $\lambda_2=0$ (green line), or derive $\lambda_2$ from requiring correct global minimum (in orange) or absence of false vacua (in blue). Dots (from darker to lighter) are derived from numerical scans with $n_\chi=2,3,6$ respectively, requiring absence of false vacua in scalar potential.}\label{fig:alpha_lambda12}}
\end{figure}

Let us now derive an analytic lower bound on $\alpha_{\text{SNR}}$, and hence on the number of new fermions. 
To minimize the right-hand side of Eq.~(\ref{eq:alphasnr}) we will set the second Higgs mass close to the minimal phenomenologically allowed value $m_{h2}=63$~GeV and, at first, also assume negligible quartic coupling $\lambda_2$. The corresponding dependence of the minimal $\alpha_{\text{SNR}}$ on $\lambda_{12}$ is shown in Figure~\ref{fig:alpha_lambda12} as the green line. This bound is obtained by imposing Eq.~(\ref{eq:alphasnr}) at $T=130$~GeV and 500~GeV, the lower-$T$ part is dominated by $1/T^2$ terms, while the higher-$T$ part is dominated by constant terms. We omit the $\propto T^2$ contribution of Eq.~(\ref{eq:alphasnr}) whose effect was already estimated in the $n_\chi$-independent bound of Eq.~(\ref{eq:tsnrmax1}). Also, at $T=130$~GeV we assume the bound to be twice what is actually obtained in high-$T$ expansion Eq.~(\ref{eq:alphasnr}) in order to account for the suppression of the fermionic contribution to the Higgs potential at low $T$, as was discussed after Eq.~(\ref{Vfull}). 

A more refined bound on $\alpha_{\text{SNR}}$ can be obtained by noticing that at negative $m_2^2$ the $h_2$ quartic has to be positive to ensure stability of scalar potential. Imposing a lower bound on $\lambda_2$ derived from the correct global minimum condition Eq.~(\ref{eq:correctglobalmin}) and from the absence of false vacuum Eq.~(\ref{eq:secondlocmin}), we obtain respectively the orange and blue curves in Figure~\ref{fig:alpha_lambda12}. Gray dots of different shades show the results of numerical scans  for $n_\chi=2,3,6$ (more details are given in Section \ref{sec:numscan}) with a requirement to have no false vacua. 
Using the analytic results we obtain
\begin{eqnarray}\label{eq:alphasnrmin}
\alpha_{\text{SNR}}&\gtrsim& 0.3.
\end{eqnarray}
The fact that the numerically obtained values of $\alpha_{\text{SNR}}$ somewhat exceed the analytically derived lower bound is explained by the effect of the terms $\propto T^2$ in Eq.~(\ref{eq:alphasnr}) that we neglected. Furthermore, the bound on $\alpha_{\text{SNR}}$ can be recast as a bound on $n_\chi$
\begin{eqnarray}\label{eq:minnchi}
n_\chi= \frac{\alpha_{\text{SNR}} \Lambda}{m_{\chi 0} }\gtrsim 2,
\end{eqnarray}
where we have assumed that $\Lambda\geq1$~TeV and $m_{\chi 0}\leq200$~GeV. As we will show in the next subsection, SNR can be achieved for even higher $m_{\chi 0}$, but an increase of $m_{\chi 0}$ would need to be accounted for by a further suppression of the fermionic contribution to the Higgs mass at low temperature compared to the high-$T$ expansion result. The estimate in Eq.~(\ref{eq:minnchi}) is confirmed by our numerical parameter space scan.

As the temperature drops further, the Higgs fields will eventually transit to the $h_1\ne 0$, $h_2=0$ vacuum. Once this happens, the electroweak symmetry breaking will be supported by the first Higgs doublet which alone provides $h/T>1$ for $T\lesssim 130$~GeV. 
In the presence of the shallower minimum at $h_1=0$ the system can remain in false vacuum for some time and then tunnel. In some cases the transition can be smooth and go through a global minimum with both Higgses having non-vanishing VEVs.

\subsection{Numerical Scan} \label{sec:numscan}

\begin{figure}[t]
\begin{center}
\includegraphics[width=145pt]{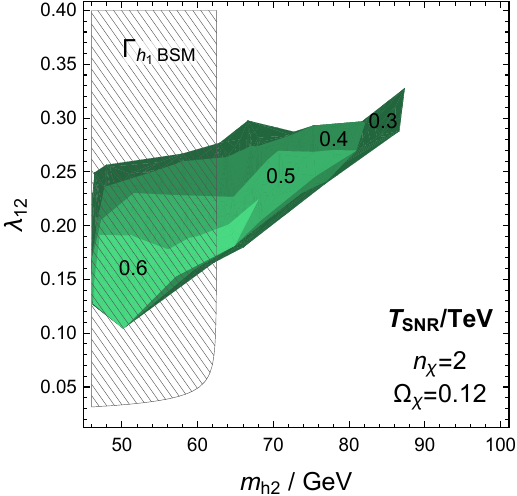}
\includegraphics[width=145pt]{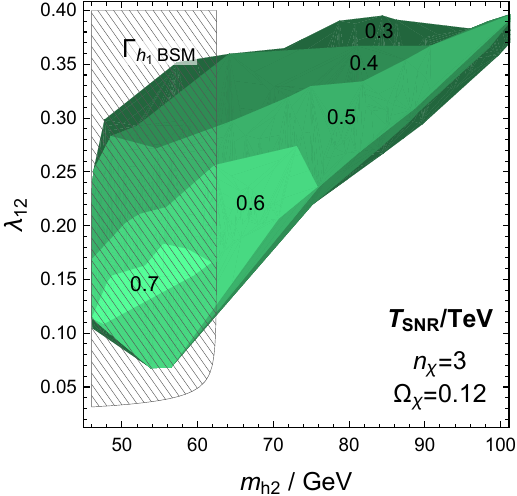}
\includegraphics[width=145pt]{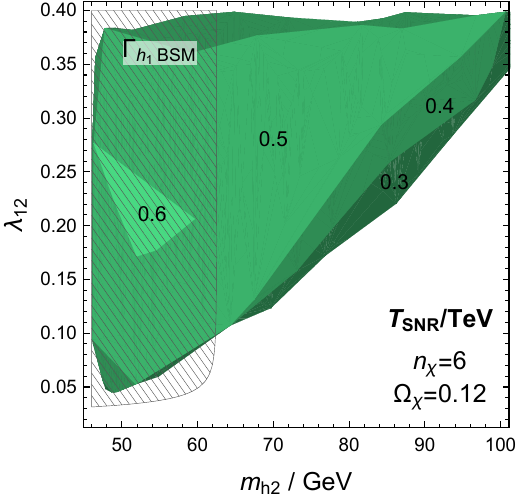}
\vspace{-7mm}
\caption{\small \it{Contours of maximal SNR temperature (in TeV), as a function of the physical $h_2$ mass and the cross-quartic $\lambda_{12}$. Meshed area is excluded by the BSM Higgs width bound.
\label{fig:scatplots1}}}
\end{center}
\end{figure}

In this section we  present the results of the numerical scans of the model parameter space in Figure~\ref{fig:scatplots1} and Figure~\ref{fig:scatplots2} for $n_\chi=2,3,6$. The details of the potential are presented in Appendices~\ref{app:tcorr}~\&~\ref{app:veff}. The parameters  are scanned within the following ranges:
$\lambda_{12}=[0,0.4]$, $\lambda_{2}=[10^{-4},0.3]$, $m_{h2}=[46,150]$~GeV, $m_{\chi 0}=[150,800]$~GeV, and $\Lambda=[1,3]$~TeV. The scale $\tilde \Lambda$ is fixed to reproduce the DM relic abundance with $\tilde \Lambda \simeq 6$~TeV, for $n_\chi=2$, $\tilde \Lambda \simeq 4.5$~TeV, for $n_\chi=3$ and $\tilde \Lambda \simeq 2.7$~TeV for $n_\chi=6$ (see Appendix~\ref{app:relic}). Only points with a single minimum of the scalar potential within $h_{1,2} < \Lambda,\tilde \Lambda$ are shown.\footnote{Models with two minima are  numerically more complicated, as one needs to track the metastable vacua. The tunneling action at each temperature must be computed to ascertain the lifetime of the metastable vacuum to check if it is phenomenologically acceptable. To simplify the analysis we rejected all points with metastable minima, thus our scans are conservative and potentially the viable parameter space could be somewhat larger.}

Figure~\ref{fig:scatplots1} shows the dependence of the maximal SNR temperature on $m_{h2}$ and $\lambda_{12}$, which are the variables most relevant when considering the collider tests of our scenario. Figure~\ref{fig:scatplots2} shows the maximal SNR temperature as function of $m_\chi$ and $\lambda_{12}/\Lambda$, which are the parameters relevant for DM direct detection experiments. In  Section \ref{sec:expbounds} we will discuss the experimental tests in more detail. As for the remaining parameters not shown in the plots, successful SNR requires $\Lambda \lesssim 1.5$~TeV (with SNR temperatures maximized at maximal $\Lambda$) and $\lambda_{2} \lesssim 0.1-0.2$.

\begin{figure}[t]
\begin{center}
\includegraphics[width=145pt]{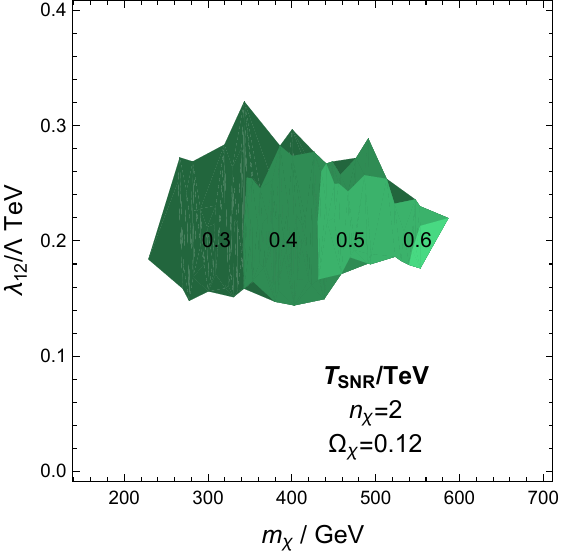}
\includegraphics[width=145pt]{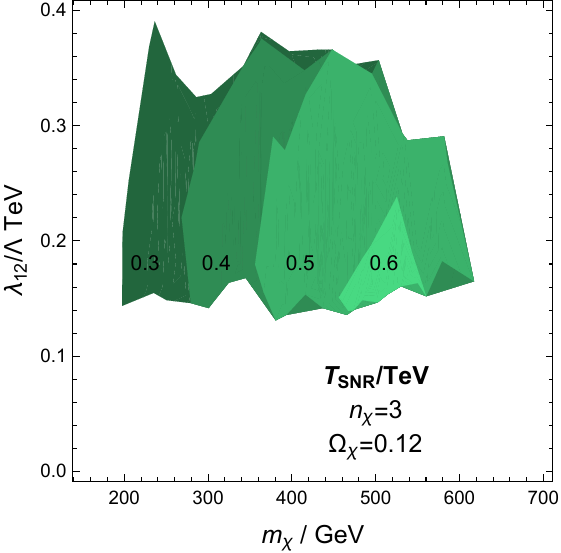}
\includegraphics[width=145pt]{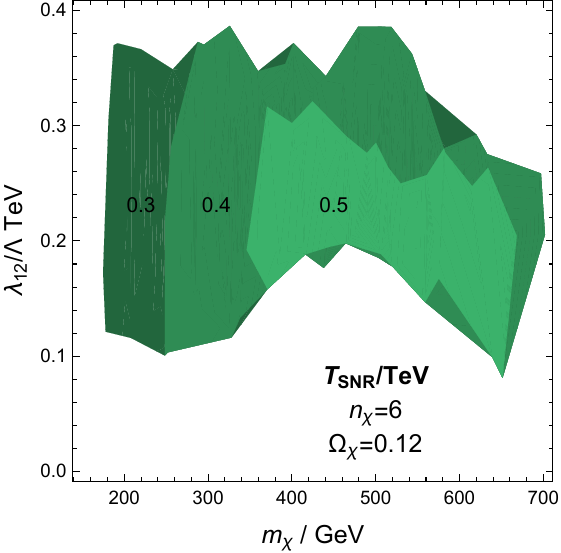}
\vspace{-7mm}
\caption{\small \it{Contours of maximal SNR temperature (in TeV), as a function of the $\chi$ mass and $\lambda_{12}/\Lambda$. Colored regions satisfy the collider constraint on the $h_1$ width, which mostly limits them from below.
\label{fig:scatplots2}
}}
\end{center}
\end{figure}

We first note that SNR with $n_\chi=1$ requires a rather low scale $\Lambda$, around $0.6$~TeV, and therefore does not allow to reproduce the DM density. Having performed a separate scan with correspondingly low $\Lambda$, we found the maximal SNR temperatures in this case to be $\sim450$~GeV at $\lambda_{12}\simeq 0.18$ and $m_{h2} \simeq 63$~GeV.  We comment further on this case in Section \ref{sec:remarks}.

For $n_\chi>6$ the maximal SNR temperatures drop (this is the effect of decreasing $\tilde \Lambda$ as we discuss below), while $\alpha_{\text{SNR}}$ increases leading to the overall parameter space region with SNR growing. The estimate of minimal $\alpha_{\text{SNR}}$ of Eq.~(\ref{eq:alphasnrmin}) together with the perturbativity bound  Eq.~(\ref{eq:pert}) predicts the growth of the maximal SNR temperature with $n_\chi$ as follows
\begin{eqnarray}
T \propto \sqrt{n_\chi} m_{\chi 0}\,,
\end{eqnarray}
however this deviates from what we find numerically. The reason for this is that once we fix the DM density via Eq.~(\ref{eq:relic1hps}), the scale of the pseudo-scalar interaction becomes tied to the number of new fermions $\tilde \Lambda \simeq 6\,\text{TeV}/\sqrt{n_\chi}$. At larger $n_\chi$ the scale $\tilde \Lambda$ decreases, leading to a larger one-loop contribution to the Higgs potential, which can eventually produce a barrier within $h_2<\Lambda$, followed by a run-away region. Parameter space points with correspondingly large $\Lambda$, which cannot be reliably described within our EFT, are therefore not considered in our scan. Discarding large $\Lambda$ in turn implies lower SNR temperatures, subject to the perturbativity bound $T\lesssim\Lambda/\sqrt{n_\chi}$. Specifically, we find that the SNR temperatures of at least $0.5$~TeV can only be achieved with $n_\chi \leq 9$.
While there might be ways to deal with the run-away problem (e.g.~arranging for appropriate higher-dimensional operators or using some specific UV-completion), another perturbativity bound $T\lesssim\tilde \Lambda/\sqrt{n_\chi}$ combined with the condition for the relic density $\tilde \Lambda \simeq 6\,\text{TeV}/\sqrt{n_\chi}$ also eventually limits the number of new fermions to at most 10 for SNR temperatures in excess of $500$~GeV.

Importantly, the results derived for $n_\chi\leq3$ are not affected by the mentioned perturbativity bounds or the runaway behavior of the Higgs potential, which allows to consider them as a more robust prediction of our EFT.

\section{Experimental bounds} \label{sec:expbounds}

Having identified the parameter space for which electroweak SNR can be achieved while simultaneously reproducing the DM relic abundance, we next examine the experimental constraints on these models, as well as the future detection prospects. Specifically, we examine the constraints coming from DM direct detection and collider bounds on exotic Higgs decays. 

We do not explore indirect detection bounds as these are expected to be model dependent and subleading to those from direct detection over the parameter space of interest. The dominant annihilate channel is $\overline{\chi}\chi\rightarrow h_2h_2$ and with the $h_2$ decaying to fermions via their small Yukawa couplings (which could differ from the SM Higgs). This introduces a degree of model dependence in the indirect detection bounds. Assuming, the $h_2$ have modest couplings to $b$-quarks such that $\overline{b}b$-pairs are the dominant annihilation product, we find that the annihilation cross-section is about 3 orders of magnitude below the current sensitivity of Fermi-LAT~\cite{Fermi-LAT:2015att}.

\subsection{DM Direct Detection}

\begin{figure}[t]
\begin{center}
\includegraphics[width=90pt]{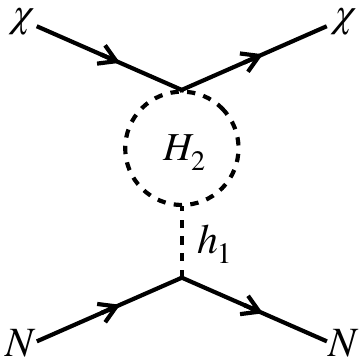}
\end{center}
\caption{\small \it{Dominant contribution to the $\chi$-nucleon scattering.}\label{fig:DDmain}}
\end{figure}

The bounds from DM direct detection experiments were fatal to the viability of SNR fermions as DM candidates in the original (single Higgs boson) SNR scenarios. In the model extended with the second Higgs doublet these bounds become significantly relaxed: in the absence of  a $H_2$ VEV today and vanishing $H_2$ couplings to SM quarks, the direct detection cross-section becomes one-loop suppressed (see Figure~\ref{fig:DDmain} for the diagram of the dominant contribution). Moreover, the one-loop interaction between nucleons and $\chi$ has to proceed via the cross-quartic $\lambda_{12}$ which is also typically less than $1$. The leading one loop diagram generates the following effective interactions between quarks and $\chi$ particles
\begin{equation}
{\cal L} \supset 
\sum_q a_q [\overline q q][\overline \chi \left(1  +  {i \gamma_5} \frac{\Lambda} {\tilde \Lambda}\right) \chi],
\end{equation} 
where the parametric estimate (see Appendix~\ref{app:dd}) of the effective couplings $a_q$ reads 
\begin{equation}\label{eq:aq_main}
 a_q = \frac{\lambda_{12}}{4 \pi^2} \frac{m_q}{m_{h1}^2 \Lambda} \log \frac{\Lambda^2}{m_{h2}^2}.
\end{equation}

While the overall quark-$\chi$ coupling is also expected to receive UV contributions from the cutoff physics,  Eq.~(\ref{eq:aq_main}) can be used as a lower bound (assuming no cancellations happen between different contributions). 
The resulting DM-nucleon cross section is~\cite{Agrawal:2010fh, Tsai:2013bt}
\begin{align}\label{eq:xs1loop}
  \sigma_{\chi N}^{\rm (1-loop)}&=\frac{\mu_{\chi N}^{2}}{\pi} \frac{1}{A^{2}} \left[Z g_{p}+(A-Z)g_{n})\right]^{2},
\end{align}
where $\mu_{\chi N}$ is the reduced mass of DM and a nucleon, $Z$ and $A$ are the atomic and mass numbers of the target nuclei, and we have neglected the subleading contribution due to the pseudo-scalar interactions. The effective DM-nucleon couplings $g_{N}$ are defined as
\begin{align}
g_N=\sum_q a_q \langle N|\overline q q|N\rangle,
\end{align}
with the matrix elements of quark currents in a nucleon $N=\{p,n\}$ given by~\cite{Goodman:1984dc,Shifman:1978zn}
\begin{eqnarray}
\langle N | \overline q q |N \rangle = 
\begin{cases} 
q=u,d,s &\quad \frac{m_N}{m_q} f_{Tq}^{(N)} \\[5pt]
q=c,b,t & \quad \frac 2 {27} \frac{m_N}{m_q} (1-\sum_{q=u,d,s}f_{Tq}^{(N)})
\end{cases}.
\end{eqnarray}
The values of $f^{(n)}_{Tq}$ can be computed in chiral perturbation theory using measurements of pion-nucleon sigma term~\cite{Cheng:1988cz,Cheng:1988im,Alarcon:2011zs,Gasser:1990ce}, where we use a representative set of values adopted in the DarkSUSY package~\cite{Gondolo:2004sc}
\beq
  f^{p}_{T u}=0.023,~~~~f^{p}_{Td}=0.034,~~~~f^{p}_{Ts}=0.14,\\
  f^{n}_{T u}=0.019,~~~~f^{n}_{Td}=0.041,~~~~f^{n}_{Ts}=0.14.
\eeq

\begin{figure}[t]
\begin{center}
\includegraphics[width=310pt]{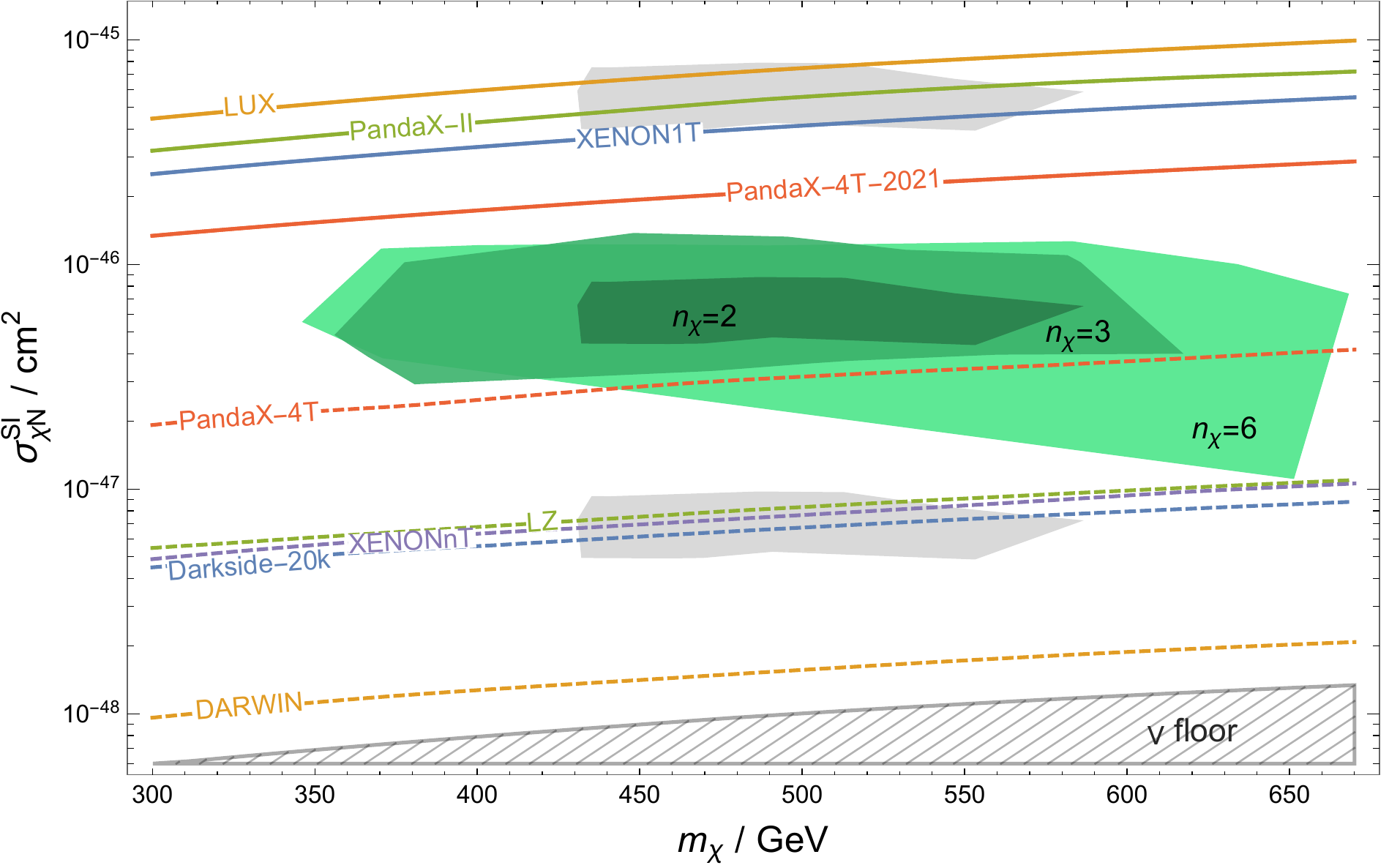}
\vspace{-4mm}
\caption{\small \it{Current (XENON1T \cite{Aprile:2018dbl}, LUX \cite{Akerib:2016vxi}, PandaX-II \cite{Cui:2017nnn}, PandaX-4T-Commissioning Run \cite{PandaX-4T:2021bab}) and projected (XENONnT \cite{Aprile:2020vtw}, DARWIN \cite{Aalbers:2016jon}, LZ \cite{Akerib:2018lyp}, PandaX-4T \cite{Zhang:2018xdp}, Darkside-20k \cite{Aalseth:2017fik}) experimental bounds on spin-independent nucleon-DM cross-sections as a function of DM mass.
The green regions indicate the cross-sections corresponding to SNR with $T\geq 0.5$~TeV for $n_\chi=2,3,6$ in the parameter range satisfying the Higgs physics bounds, for the $\chi$-quark couplings given in Eq.~(\ref{eq:aq_main})
The gray areas are obtained by increasing or decreasing the $\chi$-quark couplings by a factor of 3, for $n_\chi=2$. The neutrino floor is also displayed. 
\label{fig:ddplots}}}
\end{center}
\end{figure}

Figure~\ref{fig:ddplots} shows that the DM-nucleon cross sections favored by SNR lay in a region between the  excluded area and the reach of currently constructed experiments.
Note that even a factor of 3 enhancement (or suppression) of $a_q$, which in fact are free parameters in our EFT, can bring a large fraction of the SNR-favored parameter space into the excluded region (or outside the region that will be probed in near future experiments). 
Notably, in case of such a coupling suppression most of the green region will remain above the neutrino floor.

\subsection{Collider Bounds}
\label{sec:collider}

If $m_{h1}>2\,m_{h2}$ the Higgs boson can decay into a pair of $H_2$ components, with the total decay width given by
\begin{align}
	\Gamma_{h_1\to H_2 H_2}=\frac{\lambda_{12}^{2}v^{2}}{8\pi \, m_{h_{1}}}\sqrt{1- \frac{4m_{h2}^2}{m_{h1}^{2}}},
\end{align}
which, depending on the $H_2$ decay mechanism can be classified as invisible or untagged $h_1$ decays~\cite{Fuchs:2020cmm}. Such exotic Higgs decays into $H_2$ are bounded by $\text{BR}_{\text{BSM}}<0.2$~\cite{Bechtle:2014ewa}. Neglecting the phase space suppression factor, this results in a bound $\lambda_{12} \lesssim 0.007$.

In the opposite case $m_{h1} < 2\,m_{h2}$ the strongest bound on $\lambda_{12}$ was derived from a modification of the 1-loop $h_1$-photon coupling~\cite{Carena:2021onl}  (assuming $m_{h2}=m_{h1}/2$) and reads $\lambda_{12} \lesssim 0.4.$ This bound will potential be improved by about one order of magnitude at HL-LHC~\cite{Cepeda:2019klc}. 
Both discussed current bounds were imposed in the plots presented in Figures~\ref{fig:scatplots1} and~\ref{fig:scatplots2}.

\section{Concluding Remarks}\label{sec:remarks}

This paper has highlighted that DM can extend electroweak symmetry breaking to higher temperatures. The main ingredient allowing for the link between DM and SNR is the addition of a second Higgs doublet, which both facilitates the symmetry breaking and suppresses  direct detection signals. Unlike scenarios in which SNR is not linked to DM, our model has an upper bound on the number of SNR states: $n_\chi \lesssim 10$. High-temperature symmetry breaking at this low number of new states relies on the non-vanishing positive cross-quartic between the two Higgs doublets and requires the second Higgs to be relatively light. These facts imply that the model will be tested at the forthcoming collider and, potentially, direct detection experiments. The latter can, however, be evaded with a mild tuning of the DM-quark interactions (with the exception of DARWIN). While a positive identification at a direct detection experiment would not be sufficient to confirm this model, however it is notable that the model has implications for collider signatures which would allow for correlated signals. In particular, one might be optimistic regarding probing the second Higgs state $H_2$ through precision Higgs studies.

The model studied here features an explicit cutoff $\Lambda$, which calls for a UV completion. The desired $\overline{\chi}\chi |H_2|^2$ interactions can be produced in multiple ways, for instance, by integrating out a heavy scalar with a Yukawa coupling to $\chi$ fermions, integrating out new $SU(2)_L$ doublet fermions, or can be produced by some new strong dynamics at the scale $\Lambda$~\cite{Matsedonskyi:2020mlz}.  Since the cutoff is necessarily TeV-scale this implies that the UV physics can be potentially probed in the near future collider experiments, and may be linked to the hierarchy problem (as in \cite{Matsedonskyi:2020kuy}).

Notably, the fact that SNR is achieved with a small number of additional states is very elegant (as opposed to $\mathcal{O}$(100) in the original bosonic models \cite{Meade:2018saz,Baldes:2018nel,Glioti:2018roy}).  Indeed, it may be tempting to draw a connection between the three generations of the Standard Model and the $n_\chi=3$ SNR model, which is seen to work well in Figure \ref{fig:ddplots}. The first detailed analysis of the effect of the second Higgs doublet on SNR was recently performed in \cite{Carena:2021onl} for the case of SNR with new singlet scalars, this work also predicting a lower number  of SNR scalars (order-20) compared to the original single Higgs models  \cite{Meade:2018saz,Baldes:2018nel,Glioti:2018roy}. Moreover, these authors presented a detailed phenomenological analysis regarding detection prospects of the second Higgs doublet, which is highly relevant to the models outlined in this work.

Let us also comment further here on whether the case of $n_\chi=1$ can be accommodated within our scenario. Setting $n_\chi=1$ implies that the SNR condition Eq.~(\ref{eq:alphasnrmin}) leads to an upper bound: $\Lambda \lesssim 0.7$~TeV (close to the value obtained in our numerical scan). While such values of the cutoff appear dangerously close to the temperatures at which we would like SNR to operate, we now also encounter a problem of DM underproduction. With such low $n_\chi$ and $\Lambda$, the scalar interaction alone leads to the $\chi$ relic density $\Omega_\chi \lesssim 0.05$. Thus, while $n_\chi=1$ is interesting for SNR (provided that the low cutoff does not interfere with SNR, which can be checked in a specific UV completion), it cannot explain the observed value $\Omega_{\text{DM}} \simeq 0.12$.

Small changes to our model give alternative scenarios which may also be of interest. For instance, new SNR fermions might be involved in the generation of neutrino masses, however, in this case it would seem unlikely that they could also play the role of DM. Alternatively, the neutral component of the second Higgs doublet could also be stable, while the sizable cross-quartic required by SNR with low $n_\chi$ implies that  this state cannot account for all of the DM without tension with direct detection  limits \cite{Kalinowski:2019cxe}, giving a motivated route to a two-component DM model. We leave these variations  for future studies. 

As alluded to in the introduction, the principle motivation for delaying electroweak symmetry restoration until high temperatures is the realisation of electroweak baryogenesis at higher temperatures. Since the analysis of the strength of the phase transition and baryon number violation are distinct sets of calculations, which also require additional model building, we have chosen to study the connections between electroweak symmetry non-restoration and dark matter in isolation from these other issues. Implications of such high temperature electroweak baryogenesis have been explored in \cite{Baldes:2018nel,Glioti:2018roy,Matsedonskyi:2020mlz}. However, the scenario that we have studied here is distinct from these previous models since in order to reduce the number of $\chi$ states in our theory and satisfy direct detection bounds, a second Higgs field $H_2$ was introduced which significantly alters  the vacuum structure of the theory.

 Notably, at temperatures above $\sim$100 GeV it is $H_2$ which is the primary source of electroweak symmetry breaking. Thus the universe essentially undergoes a two-step phase transition evolving from the unbroken phase, through a phase in which the $H_2$  VEV breaks the electroweak symmetry, and then finally transitioning to the observed low energy vacuum state with the Standard Model Higgs being the main source of  electroweak symmetry breaking. Successful mechanisms of baryogenesis via (low-temperature) two-step electroweak phase transitions have been previously explored in the literature  \cite{Patel:2012pi,Blinov:2015sna,Inoue:2015pza}, and the EW phase transition can be made first order with appropriate parameter choices. Building on the results of this work, we plan to report on the prospects for realisation of electroweak baryogenesis in the context of this model in a future publication.

\appendix

\section{Thermal Corrections} \label{app:tcorr}

At finite temperature $T$ the scalar potential receives the following corrections
\beq\label{eq:1loopvh}
\Delta V_b^T = \frac{T^4}{2\pi^2} J_b[{m^2}/{T^2}],\qquad
\Delta V_f^T = -\frac{2T^4}{\pi^2} J_f[{m^2}/{T^2}]
\eeq
respectively for one thermalized bosonic degree of freedom and one Dirac fermion with mass $m$. Their interactions with the Higgs field are encoded in the Higgs-dependent masses $m$. The thermal loop functions are defined as
\beq
J_b[x]= \int_0^\infty dk\, k^2 \log \left[1- e^{-\sqrt{k^2+x}} \right],\qquad
J_f[x]= \int_0^\infty dk\, k^2 \log \left[1+ e^{-\sqrt{k^2+x}} \right].
\eeq 
In the high-temperature limit $m^2/T^2\ll 1$ they simplify to
\beq\label{eq:Vbfexp}
\Delta V_b^T \simeq -\frac{\pi^2 T^4}{90}  + \frac{T^2 m^2}{24} ,\qquad
\Delta V_f^T \simeq -\frac{7\pi^2 T^4}{180}  + \frac{T^2 m^2}{12}.
\eeq

Let us consider the correction induced by the new fermions with mass $m=m_{\chi0}-h^2/\Lambda$. Applying the high-$T$ expansion, we find the Higgs mass correction grows with $m_{\chi 0}$ 
\beq\label{eq:deltamh}
\delta m_h^2 \simeq - T^2 \frac{m_{\chi 0}}{3 \Lambda}\,,
\eeq
while the true quadratic correction is given by
\beq
\delta m_h^2 \simeq \frac{8 T^3}{\pi^2 \Lambda} (m_{\chi 0}/T) J_F'[(m_{\chi 0}/T)^2] \,,
\eeq
and its absolute size is maximized at $m_\chi^0/T \simeq 1.4$ where it reaches $\sim 1/2$ of the naively expected value in Eq.~(\ref{eq:deltamh})
\beq
\delta m_h^2 [m_{\chi 0}\simeq 1.4\, T] \simeq - 0.7\, \frac{T^3}{3 \Lambda} \simeq - 0.5\, T^2 \frac{m_{\chi0}}{3 \Lambda}.
\eeq

The overall thermal correction to the Higgs potential is given by
\begin{eqnarray}\label{eq:vhtfullsm}
\delta V(T) &=& 
\sum_i g_i \,\Delta V_i^T + \frac{T}{12 \pi} \sum_j \kappa_j \left\{(m_j^2)^{3/2}-(m_j^2(T))^{3/2}\right\}
\end{eqnarray}
where $g_{\{h_1,G_1,h_2,G_2,W,Z,t,\chi\}}=\{1,3,1,3,6,3,3,n_\chi\}$ and $\kappa_{\{\gamma,Z,W\}}=\{1,1,2\}$. The second sum in Eq.~(\ref{eq:vhtfullsm}) corresponds to the resummed daisy diagrams with thee zero modes of the longitudinal gauge boson components, with $m_j^2(T)$ being the thermally corrected masses~\cite{Delaunay:2007wb}.

\section{Relic Abundance} \label{app:relic}

In this appendix we present the annihilation cross-sections and DM relic density for $n_\chi$ Dirac DM fermions $\chi$ which annihilate into $n_\phi$ scalars $\phi$, with the Lagrangian
\begin{align}
	\mathcal{L}\supset&  - \frac{1}{2}m_\phi^{2}\phi^{2} - m_{\chi 0}\overline\chi\chi +\overline\chi\left[\frac{1}{\Lambda}+\frac{i\gamma^{5}}{\tilde\Lambda}\right]\chi \phi^{2}.
\end{align}
The annihilation cross-section is given by
\beq
\sigma v=\frac{n_\phi}{8\pi}\sqrt{1- \frac{4m_\phi^{2}}{s}} \left(\frac{1}{\Lambda^{2}}\left(1- \frac{4m_{\chi}^{2}}{s}\right)+\frac{1}{\tilde\Lambda^{2}}\right),
\eeq
where the physical $\chi$ mass is
\beq
m_\chi^2 = \left(m_{\chi 0} - \frac{\phi^2}{\Lambda}\right)^2 +  \frac{\phi^4}{\tilde \Lambda^2}.
\eeq
The annihilation cross-section can be decomposed into $s$ and $p$ wave components as
\beq	
\sigma v = a + b v^2,
\eeq
where $v$ is the relative velocity between two $\chi$ particles in the centre of mass frame and
\begin{eqnarray}
a&=& \frac{n_\phi}{8\pi\tilde\Lambda^{2}}\sqrt{\frac{m_{\chi}^{2}-m_{\phi}^{2}}{m_{\chi}^{2}}}, \\
b&=& \frac{n_\phi m_{\phi}^{2}}{64\pi\tilde\Lambda^{2}(m_{\chi}^{2}-m_{\phi}^{2})}\sqrt{\frac{m_{\chi}^{2}-m_{\phi}^{2}}{m_{\chi}^{2}}}+\frac{n_\phi}{32\pi\Lambda^{2}}\sqrt{\frac{m_{\chi}^{2}-m_{\phi}^{2}}{m_{\chi}^{2}}}.
\end{eqnarray}
It follows that the relic abundance of $\chi$ is given by~\cite{Bai:2013iqa}
\begin{align}
	\Omega h^{2}&\simeq \frac{1.07 \times 10^{9}}{\text{GeV} M_{\text{Pl}}\sqrt{g_{*}}}\frac{x_{F}}{a+3(b-a/4)/x_{F}}\times 2n_{\chi},
\end{align}
where $M_{\text{Pl}}\simeq 1.22 \times 10^{19}$~GeV is the Planck mass. 
In our numerical calculations we take $g_* \simeq 86$ and $x_F\simeq25$.

\section{Direct Detection Cross-Sections} \label{app:dd}

In this appendix we provide details on the calculation of the direct detection cross-sections used in Section~\ref{sec:expbounds}.

\begin{figure}[b!]
\centering
  \includegraphics[scale=0.4]{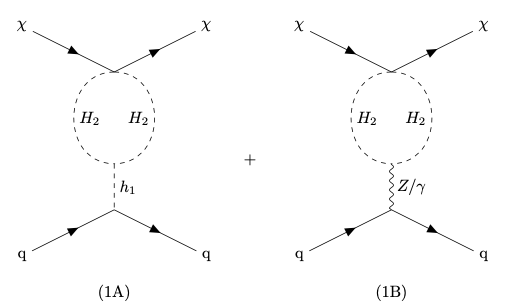}
  \vspace{-4mm}
  \caption{DM-quark interaction at one-loop level (produced using {TikZ-Feynman}~\cite{Ellis:2016jkw}). \label{feyn1}}
\end{figure}

\subsection{Tree level}

We parametrize the effective $\chi$-quark interaction obtained after integrating out heavy mediators in the following way
\begin{equation}\label{eq:leffdd}
{\cal L}_{eff} = 
\sum_q a_q [\overline q q][\overline \chi \left(1 +  {i \gamma_5} \frac{\Lambda} {\tilde \Lambda}\right) \chi],
\end{equation} 
and the corresponding scattering cross-section is given in Eq.~(\ref{eq:xs1loop}).
The following tree-level contribution to $a_q$ would arise in the SNR scenarios with a single Higgs doublet
\begin{equation}
a_q^{(tree)} = - \frac{2 m_q}{m_{h1}^2 \Lambda}. 
\end{equation}

\subsection{One loop}

The leading contribution to $\chi$-quark interactions in the discussed two Higgs doublet model with SNR arises at one loop level. The corresponding Feynman diagrams are depicted in Figure~\ref{feyn1}. The $h_1$-mediated scattering (1A) leads to the following correction to the effective four-fermion interaction of Eq.~(\ref{eq:leffdd})
\begin{equation}\label{eq:aq1l}
a_q^{(\rm 1-loop)}= \frac{\lambda_{12}}{4 \pi^2} \frac{m_q}{m_{h1}^2 \Lambda} \log \frac{\mu^2}{m_{h2}^2},
\end{equation} 
where we used dimensional regularization with the $\overline{MS}$ scheme. 
In the initial EFT containing $h_{1,2}$ the divergence appearing in the computation of $a_q$ would be absorbed in the renormalization of the operator $h_1^2 \overline \chi \chi$, which contributes to the quark-$\chi$ interactions as
\begin{equation}\label{eq:leffdd}
\delta a_q^{(\rm contact)} = c_{h_1 \chi} \frac{m_q}{m_{h1}^2 \Lambda},
\end{equation}  
where $c_{h_1 \chi}$ is a free parameter of the EFT.
Not aiming at matching our theory to some specific UV completion, we simply estimate the resulting $\chi-q$ interactions by substituting $\mu=\Lambda$ in Eq.~(\ref{eq:aq1l}) and omitting $\delta a_q^{(\rm contact)}$. 

Note that the one-loop diagram involving gauge bosons (1B) instead vanishes, as the corresponding amplitude requires to couple (pseudo)scalar and vector currents~\cite{Kopp:2009et}.

\subsection{Two loops}

\begin{figure}[t]
\centering
  \includegraphics[scale=0.37]{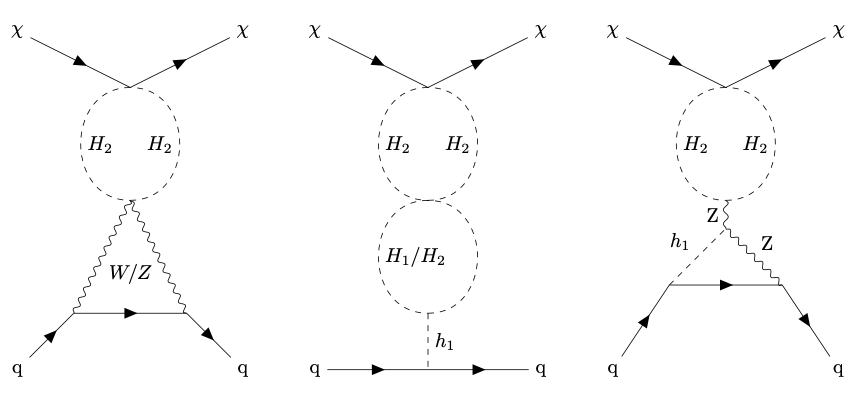}
  \caption{Representative diagrams for DM-quark interaction at two-loop level without photons.}
  \label{feyn2}
\end{figure}

The parametric estimate of the size of the two loop contributions (see Figure~\ref{feyn2}) to the four-fermion interaction is
\begin{equation}\label{eq:aq2l}
a_q^{(2-loop)}= \frac{g^4}{(4\pi)^4} \frac{m_q}{\Lambda m_W^2}, 
\end{equation} 
which represents a small correction to the 1-loop contribution with a relative size
\begin{equation}
\frac{g^4}{(4\pi)^2}/\lambda_{12} \simeq 10^{-3}/\lambda_{12}.
\end{equation}

These short-range corrections do not include the 2-loop contribution with two photon lines which has to be treated separately. Given that the latter has a different energy dependence, we will perform the comparison at the level of differential event rates. We will calculate the photon-mediated two-loop scattering in two steps by first integrating out the second Higgs doublet and then using the resulting effective operator to compute the cross-section.

\begin{figure}[t]
\centering
  \includegraphics[scale=0.5]{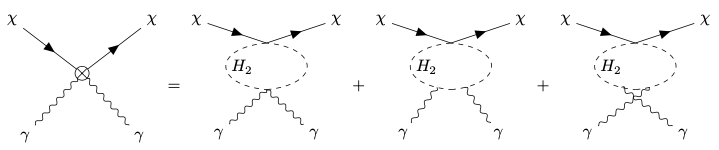}
  \caption{DM - photons interactions.  \label{feyn4}}
\end{figure}
In the first step we obtain (see Figure~\ref{feyn4})
\begin{align}
  \mathcal{L}_{\rm eff}=-\frac{e^{2}}{48\pi^{2} m_{h2}^{2}}\overline\chi  \left(\frac{1}{\Lambda}+\frac{i}{\tilde\Lambda}\gamma^{5}\right)\chi F^{\mu\nu}F_{\mu\nu},
\end{align}
which results in the following DM-nucleus scattering amplitude~\cite{Kopp:2009et} 
\begin{align}
  M=-\frac{\alpha^{2}\sqrt{2m_{i}E_{R}}}{24m_{h2}^{2}} Z^{2}\tilde F(q) \left[\overline u_{\chi}'\left(\frac{1}{\Lambda}+\frac{i\gamma_{5}}{\tilde\Lambda}\right) u_{\chi} \right]\left[\overline u^{\prime}_{N}\frac{1}{2}(1+\gamma_{0}) u_{N}\right],
\end{align}
where $E_{R}=-\frac{q^{2}}{2m_{i}}$ is the recoil energy of a nucleus, $q$ is the four-momentum transfer, $m_i$ is the target nucleus mass, $Z$ is its atomic number and $\tilde F(q)$ is the corresponding two-loop nuclear form factor. Correspondingly, the DM-target nucleus $i$ differential cross section reads
\begin{equation}\label{eq:diffxsgamma}
  \frac{d\sigma_{\chi i}^{(\gamma)}}{dE_{R}}=\frac{\alpha^{4}m_{i}^{2}Z^{4} }{576\pi \, m_{h2}^{4}v_{\chi}^{2}}E_{R} \tilde F^{2}(|q|)\left[\frac{1}{\Lambda^{2}}+\frac{1}{2}\frac{1}{\tilde\Lambda^{2}}\frac{E_{R} m_{i}}{m_{\chi}^{2}}\right]\Theta \left(v_{\chi}-\sqrt{\frac{m_{i}E_{R}}{2\mu_{\chi i}^{2}}}\right), 
\end{equation}
where $v_{\chi}$ is the DM velocity in the lab frame and $\mu_{\chi i}$ is the reduced mass of $\chi$ and a nucleus. The minimum value for $v_{\chi}$ to induce certain nuclear recoil energy $E_R$ is $v_{\chi}^{\rm min}=\sqrt{m_{i}E_{R}/2\mu^{2}_{\chi i}}$, which explains the presence of the step function.

\subsection{Event rates}\label{app:EventRate}

The differential event rate per unit target mass and per unit time is given by~\cite{Lin:2019uvt,Schumann:2019eaa}
\begin{align}
  \frac{dR}{dE_{R}}=N_i \frac{\rho_{\chi}}{m_{\chi}}\int \frac{d\sigma_{\chi i}}{dE_{R}}v_{\chi} f(\vec v_{\chi})d^3v_{\chi},
\end{align}
where $N_{i}$ is number of target nuclei per unit target mass, $f(\vec v_\chi)$ is the normalized DM velocity distribution in the lab frame, and $\rho_{\chi}=0.3$ GeV/cm$^{3}$ is the local DM density. 

The differential cross-section ${d\sigma_{\chi i}}/{dE_{R}}$ for the photon-mediated scattering was given in Eq.~(\ref{eq:diffxsgamma}) whereas for the one-loop Eq.~(\ref{eq:aq1l}) and the remaining two-loop Eq.~(\ref{eq:aq2l}) contributions it reads
\begin{align}
  \frac{d\sigma_{\chi i}}{dE_{R}}=\frac{\sigma_{\chi N}}{\mu_{\chi N}^{2}}\frac{m_{i}}{2v_{\chi}^{2}}A^{2}F^{2}(|q|)\Theta \left(v_{\chi}-\sqrt{\frac{m_{i}E_{R}}{2\mu_{\chi i}^{2}}}\right),\label{dsigma1}
\end{align}
 where $\sigma_{\chi N}$ was given in Eq.~(\ref{eq:xs1loop}) as a function of $a_q$ and $A$ is the atomic mass number.

\begin{figure}[b]
\centering
  \includegraphics[scale=0.4]{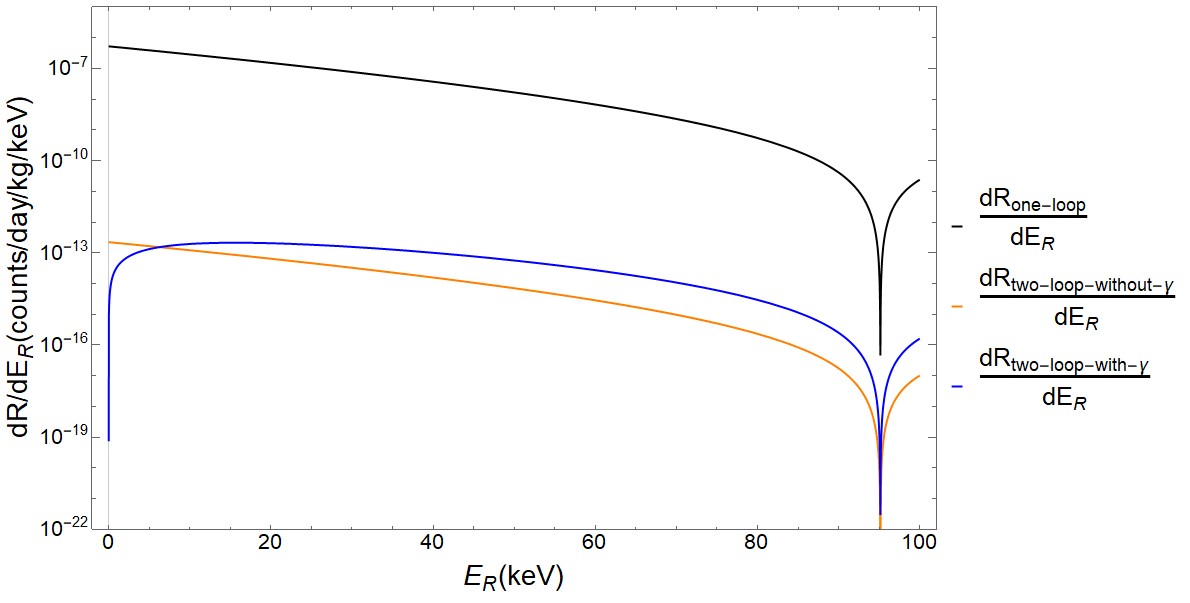}
  \caption{Comparison of the differential event rates for DM direct detection taking parameter typical values needed for SNR: $m_\chi=0.5$~TeV, $m_{h2}=70$~GeV, $\lambda_{12}=0.2$, and $\Lambda=1$~TeV.   \label{DDfig2}}
\end{figure}

To compare differential rates we take Helm form factor~\cite{Helm:1956zz} for both $F(|q|)$ and $\tilde F(|q|)$ (in the latter case the exact form factor is not expected to be important as this contribution is very suppressed), and we use the conventional truncated Maxwellian distribution \cite{Lin:2019uvt} for the  local DM velocity distribution. The comparison of differential rates for the 1-loop contribution, 2-loop contribution without photons, and the 2-loop contribution with photons is presented in Figure~\ref{DDfig2} for Xenon targets with $A=131$. For a typical SNR-favored parameter choice with $m_\chi=0.5$~TeV, $m_{h2}=70$~GeV, $\lambda_{12}=0.2$, and $\Lambda=1$~TeV the 1-loop contribution completely dominates over the others, which can therefore be safely neglected in our analysis.

\section{The $h_{1,2}$ Effective Potential} \label{app:veff}

The tree-level zero-temperature potential is fixed by Eq.~(\ref{eq:v2hdmtree}). 
The one loop zero-temperature correction to the $h_1$ and $h_2$ potential in the Landau gauge is given by
\begin{eqnarray}\label{eq:deltav1loop}
\delta V_{T=0} (h_1,h_2) &=& \sum_i g_i  \frac{(-1)^{F}}{64 \pi^2} \,m_i^4 \log \left[\frac{m_i^2}{\mu^2} \right] \\
&~&+ \Delta {m_1^2} h_1^2 + \Delta m_2^2 h_2^2 + \Delta \lambda_{1} h_1^4 + \Delta \lambda_{2} h_2^4+  \Delta \lambda_{12} h_1^2 h_2^2 + \Delta {c_6} h_2^6 + \Delta {c_8} h_2^8, \nonumber
\end{eqnarray} 
with $F=0$ for bosons and $F=1$ for fermions, $g_{\{h_1,G_1,h_2,G_2,W,Z,t,\chi\}}=\{1,3,1,3,6,3,12,4 n_\chi\}$ and $m_i$ are corresponding $h_{1,2}$-dependent tree-level masses. 

We split $H_{1,2}$ into $h_{1,2}$ (which obtain a VEV at some $T$) and three $G_{1,2}$ (with vanishing VEVs).  When presenting the phenomenological bounds in the main text we assumed that in the current minimum all $H_2$ components have the same mass and same coupling to $h_1$, thereby neglecting a small splitting induced by gauge bosons loops~\cite{Cirelli:2005uq}.  We fix the counterterms $\Delta X$ by the following conditions 
\beq\label{eq:renorm}
\partial_{h_1} V_{eff}(v_{\text{SM}},0)&=0, \\ 
\partial^2_{h_1} V_{eff}(v_{\text{SM}},0)+\Delta \Pi_{h_1}&=m_{h1}^2, \\
\partial^2_{h_2} V_{eff}(v_{\text{SM}},0)&=m_{h2}^2, \\
\partial_{h_1} \partial^2_{h_2} V_{eff}(v_{\text{SM}},0)+\Delta \Gamma_{h_1 h_2^2}&=\lambda_{12} v_{\text{SM}},
\eeq
where  $\lambda_{12}$ is defined from the $h_1 h_2^2$ coupling at $q^2 =-(0.5\,\text{GeV})^2 \simeq -(10^{-3} m_\chi)^2$ relevant for direct detection experiments, and $m_{h1}^2$ and $m_{h2}^2$ are pole masses.
In $\Delta \Pi_{h_1}$ and $\Delta \Gamma_{h_1 h_2^2}$ we account for the leading contributions to the difference between the zero momentum transfer quantities (derivatives of the effective potential) and the quantities at relevant finite momentum~\cite{Delaunay:2007wb}. The top quark and IR-divergent $H_1$ contributions to the $h_1$ self-energy difference reads~\cite{Casas:1994us}
\beq
\Delta \Pi_{h_1} 
=& \frac {3 \lambda_t^2}{8\pi^2} \left[ - 2 m_t^2 \left(Z[m_t^2/m_{h1}^2]-2\right) +\frac 1 2 m_{h1}^2 \left(\log[m_t^2/\mu^2]+Z[m_t^2/m_{h1}^2]-2\right)\right]  \\
&+ \frac{3}{128 \pi^2} \frac{g^2 m_{h1}^4}{m_W^2} \left( \pi \sqrt 3 - 8 +Z[m_G^2/m_{h1}^2] \right),
\eeq
where $m_G$ is the mass of the Goldstone bosons and 
\begin{eqnarray}
Z[x]=
\begin{cases}
x>1/4, \; 2 |1-4x|^{1/2}\arctan[|1-4x|^{-1/2}] \\
x<1/4, \; |1-4x|^{1/2} \log[(1+|1-4x|^{1/2})/(1-|1-4x|^{1/2})]
\end{cases}.
\end{eqnarray} 
Moreover, the IR-divergent contribution of $H_1$ to the $h_1 h_2^2$ coupling difference is given by 
\begin{eqnarray}
\Delta \Gamma_{h_1 h_2^2} = \frac{3}{32 \pi^2} \frac{m_{h1}^2}{v_{\text{SM}}} \left(Z[m_G^2/q^2]-2 \right).
\end{eqnarray} 
Note that $m_G^2$ vanishes at the minimum of the effective potential, but the corresponding IR-divergent pieces from $\Delta \Pi_{h_1}$, $\Delta \Gamma_{h_1 h_2^2}$ and derivatives of the effective action cancel out from the renormalization conditions Eq.~(\ref{eq:renorm}). We do not include $\Delta \Pi_{h_2}$ in Eq.~(\ref{eq:renorm}) since it does not receive contributions from the top quark or with IR-divergences.

We do not impose any renormalization condition on  $\lambda_2$ since in our study it is a mute parameter which is scanned over, and we set $\Delta \lambda_2=0$. 
The non-renormalizable interaction $h_2^2 \chi^2$ induces UV-divergent corrections to the operators $\propto h_2^6,h_2^8$.  We set corresponding counterterms $\Delta c_{6,8}$ to zero as well. 
As a result of these prescriptions, the one-loop effective action is explicitly $\mu$-dependent after renormalization, and we fix $\mu=1$~TeV.

\end{document}